\documentclass{ws-mpla}

\begin{document}

\markboth{De-hua Wen, Wei Chen, Jian-xun Hu,Liang-gang Liu }
{Frame Dragging Effect on Moment of Inertia and Radius of Gyration
of Neutron Star}

%
\catchline{}{}{}{}{}
%

\title{FRAME DRAGGING EFFECT ON MOMENT OF INERTIA AND RADIUS OF GYRATION
OF NEUTRON STAR }

\author{\footnotesize De-hua Wen$^{1,2}$}

\address{$^{1}$Department of mathematics, Zhongshan University, Guangzhou
510275, China \\
$^{2}$Department of Physics, South China University of Technology,
Guangzhou 510641, China\\
 \ wendehua@scut.edu.cn}
\author{Wei Chen}
\address{Department of Physics, Jinan University, Guangzhou 510632, China }

\author{Jian-xun Hu}

\address{Department of mathematics, Zhongshan University, Guangzhou
510275, China }

\author{Liang-gang Liu}
\address{Department of physics, Zhongshan University, Guangzhou
510275, China }

\maketitle

\pub{Received (Day Month Year)}{Revised (Day Month Year)}

\begin{abstract}
Accurate to the first order in the uniform angular velocity, the
general relativity frame dragging effect of the moments of inertia
and radii of gyration of two kinds of neutron stars are calculated
in a relativistic $\sigma-\omega$ model. The calculation shows
that the dragging effect will diminish the moments of inertia and
radii of gyration.

\end{abstract}

\ccode{PACS Nos.:  04.40.Dg;   95.30.Sf;    97.10.Kc;   97.60.Jd}

\section{Introduction}
\indent As we knows, Neutron stars may be created in the aftermath
of the gravitational collapse of the core of a massive star at the
end of its life, or may be created when the mass of a accreting
white dwarf exceed the Chandrasekhar limit. The typical character
of neutron stars may be described as following: spinning rapidly,
often making several hundred rotations per second; having a small
stature with a radius $R$ of $\sim12$km and a heavy avoirdupois
with a mass $M$ on the order of 1.5 solar masses $(M_{\bigodot})$;
and having a very compact central density as high as several times
the nuclear saturation density. Neutron stars are one of the
densest
forms of matters in the observable universe. \\
\indent One of the important global property of neutron star is
the moment of inertia $I$. Eary estimates of the moment of inertia
by the energy -loss rate from pulsars spanned a wide range of $I$
\cite{Ruderman}, and several researchers have given a lower bound
on the moment of inertia of the pulsar as \cite{Trimble,Borner}
$I\geq4-8\times10^{37}\texttt{g.cm}^{2}$. Recently,
Kalogera\cite{Kalogera} etc. also studied the bound of the moment
of inertia. In order to investigate the general relativistic frame
dragging effect on the moments of inertia of neutron stars, the
moments of inertia of the non-rotating and rotating at the Kepler
frequency neutron stars
will be studied in this work.\\
\indent Here we adopt the metric signature - + + +, G=c=1.

\section{ Moment of inertia and radius of gyration of  neutron star}
\indent In relativity, the space-time geometry of a static
spherical symmetric star can be described as
\begin{equation}
ds^{2}=-e^{2\varphi}dt^{2}+e^{2\alpha}dr^{2}+r^{2}(d\theta^{2}+\sin^{2}\theta
d\phi^{2}),
\end{equation}
and the space-time geometry of a rotating star in equilibrium is
described by a stationary and axisymmetric metric of the form
\begin{equation}
ds^{2}=-e^{2\nu}dt^{2}+e^{2\lambda}dr^{2}+e^{2\psi}(d\phi-\omega
dt)^{2}+e^{2\mu}d\theta^{2},
\end{equation}
where $\omega(r)$ is the angular velocity of the local inertial
frame and is proportional to the star's rotational frequency
$\Omega$, which is the uniform angular velocity of the star
relative to an observer at infinity.\\
 \indent From the $(t,\phi)$
component of Einstein field equations, accurate to the first order
in $\Omega$, one gets\cite{Hartle}
\begin{equation}
\frac{1}{r^{4}}\frac{d}{dr}(r^{4}j\frac{d\bar{\omega}}{dr})+\frac{4}{r}\frac{dj}{dr}\bar{\omega}=0,
\end{equation}
where$j(r)=e^{-\varphi}[1-2M_{0}(r)/r]^{\frac{1}{2}}$,
$\bar{\omega}=\Omega-\omega$, which denotes the angular velocity
of the fluid relative to the local inertial frame. The boundary
conditions are imposed as $\bar{\omega}=\bar{\omega}_{c}$ at the
center, $ \frac{d\bar{\omega}}{dr}|_{\bar{\omega}_{c}=0}$ , where
 $\bar{\omega}_{c}$ is chosen arbitrarily. Integrating eq.(3)
outward from the center of the star, one would get the function
$\bar{\omega}(r)$. Outside the star, from eq.(3) one has
$\bar{\omega}(r)=\Omega-\frac{2J}{r^{3}}$, where $J$ is the total
angular mementum of the star, which takes the form\cite{Hartle}
$J=\frac{1}{6}R^{4}\frac{d\bar{\omega}}{dr}|_{r=R}$, where $R$ is
the surface radius of the star. According to the traditional
definition, we have
\begin{equation}
J=I\Omega,
\end{equation}
where $I$ is the moment of inertia. At the neutron star's surface,
in term of eqs.(3) and (4), one can get
\begin{equation}
I=-\frac{2}{3}\int^{R}_{0}r^{3}\frac{dj}{dr}\cdot
\frac{\bar{\omega}}{\Omega}dr.
\end{equation}
From the (0,0) and (1,1) components of  Einstein field equations
to the Static Spherically Symmetric star, one
gets\cite{Chandrasekhar}
\begin{equation}
\frac{d\alpha}{dr}=\frac{1}{2r}(1-e^{2\alpha}+8\pi r^{2}\rho
e^{2\alpha}),
\end{equation}
\begin{equation}
\frac{d\varphi}{dr}=\frac{1}{2r}(e^{2\alpha}-1+8\pi r^{2}p
e^{2\alpha}).
\end{equation}
Accurate to the first order in $\Omega$, substituting eqs. (6) and
(7) into eq. (5), one has\cite{Fredman}
\begin{equation}
I=\frac{8\pi}{3}\int^{R}_{0}r^{4}e^{-\varphi}\frac{\rho+p}{\sqrt{1-\frac{2M_{0}}{r}}}\cdot
\frac{\bar{\omega}}{\Omega}dr.
\end{equation}
This is just the approximate expression of the moment of inertia
accurate to the first order in $\Omega$. To the non-rotating
neutron stars, $\frac{\bar{\omega}}{\Omega}=1 $. As a first order
approximation, it is means that there is only consideration of the
frame dragging effect and without consideration of the deformation
caused by the rotation\cite{Hartle}. According to eq.(8), we can
estimate that, to the same neutron star, the moment of inertia of
the non-rotating star will be bigger than that of the rotating
star, which will be testified in our later numerical calculation.
This is just the frame dragging effect to the moment of inertia.
Approximating in  week field, according to eqs.( 6) and (7), one
gets
\begin{equation}
\frac{dj}{dr}\approx-4\pi r\rho.
\end{equation}
In Newtonian theory, $\bar{\omega}=\Omega$, so from eq.(5), we
have
\begin{equation}
I=\frac{8\pi}{3}\int^{R}_{0}r^{4}\rho dr,
\end{equation}
this is just the calculating expression of the moment of inertia
of a spherical symmetric star in Newtonian theory.\\
\indent  Similar to the Newtonian theory, in general relativity,
we also define the radius of gyration of a spherical symmetric
star as
\begin{equation}
R_{g}={(\frac{I}{M}})^\frac{1}{2}.
\end{equation}

\section{ Numerical results and discussion}
\indent There are several models to deal with the superdense
matters, such as non-relativistic models, relativistic field
theoretical models\cite{s3,Gregory:1999pm}. To different models,
the fractions of particles in the superdense matters are
different, and then the bulk properties of superdense matters are
different, that is, the EOS of them are
different\cite{s9,Luo:2004mc}. In this work, two kind of EOS will
be employed\cite{Chen}, one is for the traditional neutron stars
(TNS), in which $n, p, e, \mu $ are the main elements; the other
is for the hyperon stars (HS), in which $n, p, e, \mu, \Lambda,
\Sigma, \Xi, \Delta $  are the main elements. The EOS will be
considered in the relativistic $\sigma-\omega$ model\cite{Serot}.
Fig.(1) shows the EOS of the THS and HS. \\
\begin{figure}[th]
\centerline{\psfig{file=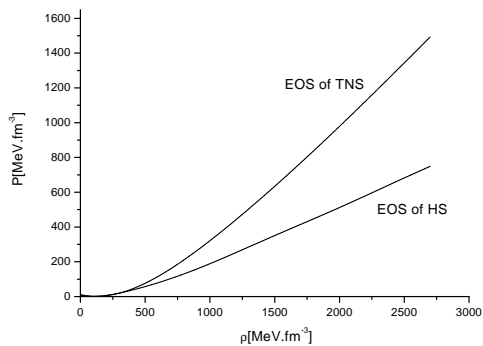,width=3.0in}} \vspace*{8pt}
\caption{Equations of state of traditional neutron stars(TNS) and
hyperon stars(HS)}
\end{figure}
\indent By using above EOS and eqs.(8) and (11), the moments of
inertia and radii of gyration of neutron stars were calculated.
Fig.(2) presents the moment of inertia of non-rotating and
rotating (at the Kepler frequency) neutron stars as a function of
central density. It is clear
 that, because of the frame dragging effect, as a first order
approximation,the moments of inertia of the non-rotating TNS (or
HS) are bigger than that of the corresponding rotating stars at
the same central density. From this figure, one can also see that,
with the same central density, as the star rotating at its Kepler
frequency, the frame dragging effect on the moment of inertia of
TNS is bigger than that of HS. The reason of this is that the EOS
 of TNS is stiffer than the EOS of HS, so there is a bigger
 Kepler frequency of TNS at the same central density, and then corresponds a bigger frame dragging effect.
In order to show the frame dragging effect of the moments of
inertia entirely, we present them in a three-dimension figure,
fig.3 just presents the moments of inertia of TNS as a function of
the central densities and the central angular velocities relative
to the local inertial
 frame. \\

\begin{figure}[th]
\centerline{\psfig{file=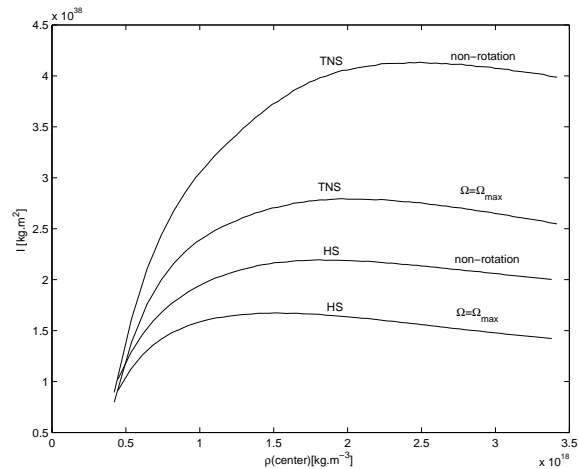,width=3.0in}} \vspace*{8pt}
\caption{The moment of inertia of non-rotating and rotating (at
the Kepler frequency) neutron stars as a function of central
density.}
\end{figure}

\begin{figure}[th]
\centerline{\psfig{file=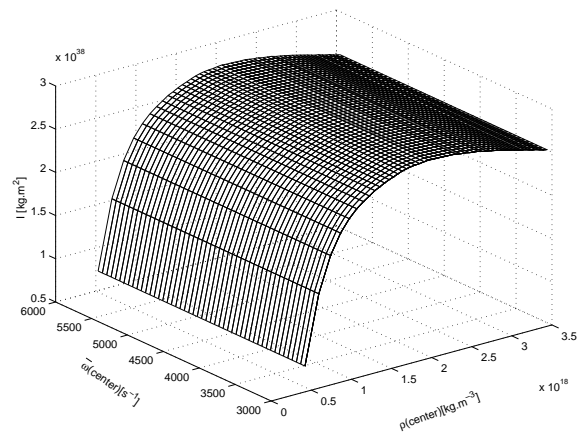,width=3.0in}} \vspace*{8pt}
\caption{The moment of inertia of TNS as a function of central
density and the angular velocity relative to the local inertial
frame at the center.}
\end{figure}

 \indent Fig.4 gives the radii of gyration of non-rotating and rotating (at
the Kepler frequency) neutron stars as a function of central
density. From this figure one can see that, to the same central
density, there is a big gap between the lines of the non-rotating
and rotating stars, that is to say, there is a remarkable frame
dragging effect to the radii of gyration, and as the central
density increasing, the effect increases. We also present the
radii of gyration of TNS
in a three-dimension figure, see fig.5.\\
\indent From our calculation, one can see that to a star at the
same central density, the frame dragging effect will diminish its
moment of inertia and radius of gyration, and as the star rotates
faster, the dragging effect will become more obviously.

\begin{figure}[th]
\centerline{\psfig{file=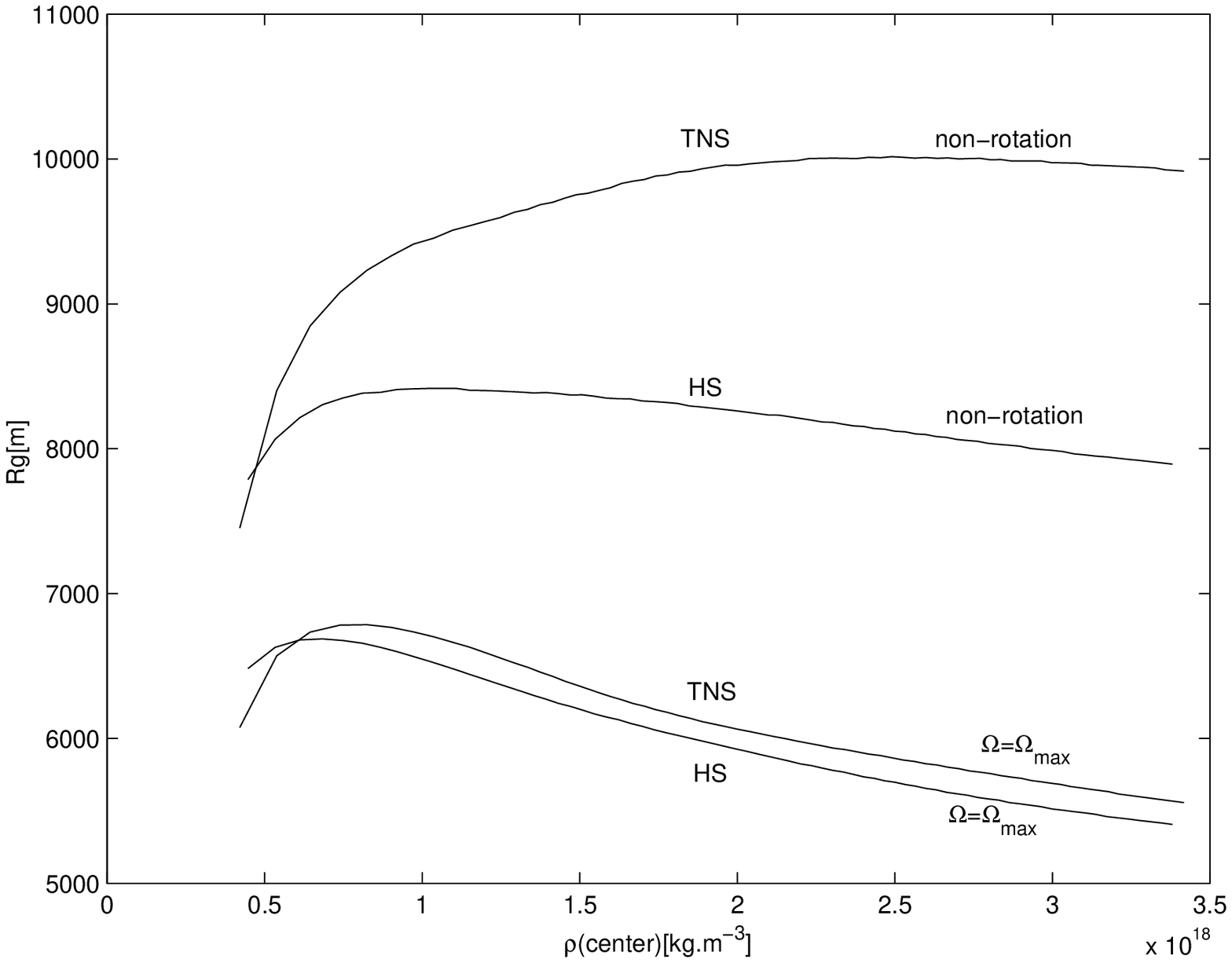,width=3.0in}} \vspace*{8pt}
\caption{The radius of gyration of non-rotating and rotating (at
the Kepler frequency) neutron stars as a function of central
density.}
\end{figure}

\begin{figure}[th]
\centerline{\psfig{file=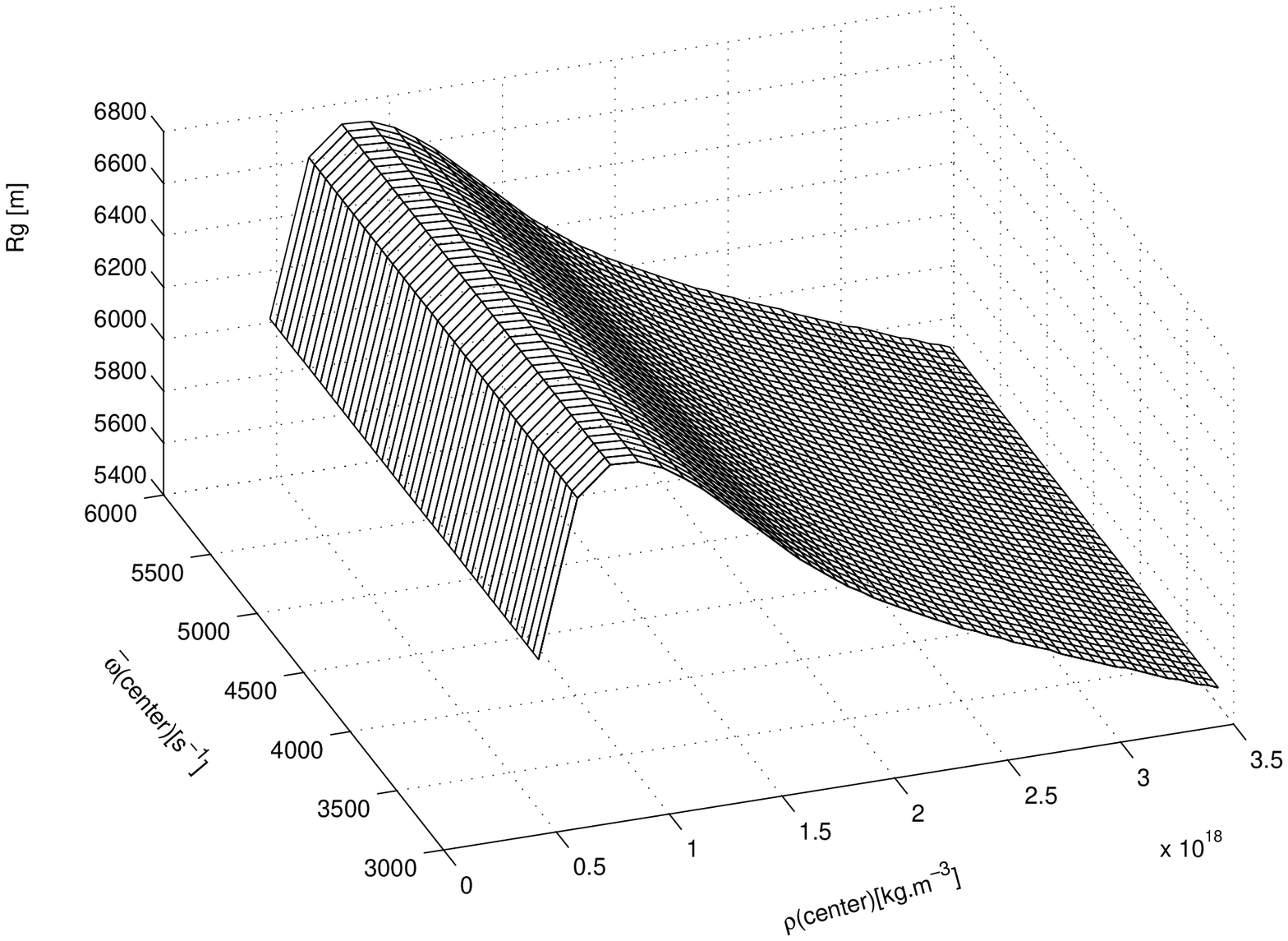,width=3.0in}} \vspace*{8pt}
\caption{The radius of gyration of TNS as a function of central
density and the angular velocity relative to the local inertial
frame at the center.}
\end{figure}

\section*{Acknowledgments}
\indent The project supported by National Natural Science
Foundation of China (Grant No. of 10275099, 10175096).\\

\end{document}